\pdfoutput=1 
\documentclass[preprint,showpacs,preprintnumbers,amsmath,amssymb,pre]{revtex4}
\usepackage[english]{babel}
\usepackage{graphicx} 
\usepackage{amssymb}
\usepackage{amsmath}
\usepackage{amsbsy}
\usepackage{dcolumn}

\usepackage[section]{placeins}
\usepackage{color}

\newcommand {\um} {~\mu \mathrm{m}}

\newcommand {\tauinf} {\tau_{\infty}}
\newcommand {\tc} {t_\mathrm{c}}

\begin{document}

\title{Plasticity of a colloidal polycrystal under cyclic shear}

\author{Elisa Tamborini}
\author{Luca Cipelletti}
\email{luca.cipelletti@univ-montp2.fr}
\author{Laurence Ramos}
\email{laurence.ramos@univ-montp2.fr}
\thanks{L. Cipelletti and L. Ramos contributed equally to this work.}

\affiliation{
Universit\'{e} Montpellier 2, Laboratoire Charles Coulomb
UMR 5221, F-34095, Montpellier,
France\\
CNRS, Laboratoire Charles Coulomb UMR 5221, F-34095, Montpellier, France\\
}

\date{\today}

\begin{abstract}

We use confocal microscopy and time-resolved light scattering to investigate plasticity in a colloidal polycrystal, following the evolution of the network of grain boundaries as the sample is submitted to thousands of shear deformation cycles. The grain boundary motion is found to be ballistic, with a velocity distribution function exhibiting non-trivial power law tails. The shear-induced dynamics initially slow down, similarly to the aging of the spontaneous dynamics in glassy materials, but eventually reach a steady state. Surprisingly, the cross-over time between the initial aging regime and the steady state decreases with increasing probed length scale, hinting at a hierarchical organization of the grain boundary dynamics.

\end{abstract}
\maketitle

The mechanical properties of amorphous solids are a topic of intense research. Recent works focus on the irreversible (or plastic) rearrangements at the microscopic level~\cite{Falk1998,Bocquet2009,LemaitrePRL2009,TsamadosEPJE2010,BarratPRL2011,Jop2012,Koumakis2012}, resulting from an applied deformation or stress, which are ultimately responsible for the macroscopic mechanical behavior. Research on amorphous solids is also relevant to crystalline materials~\cite{Biswas2013}. On the one hand, simulations and experiments have revealed that particles in the grain boundaries (GBs) separating crystalline grains exhibit glassy dynamics~\cite{Zhang2009,Nagamanasa2011}. On the other hand, polycrystals may be regarded as an amorphous assembly of crystalline grains separated by GBs. In fact, driven polycrystals display mechanical features similar to those of amorphous solids~\cite{Shiba2010} and GB process has been shown to be at the origin of the plasticity of polycrystals in the limit of small grain sizes~\cite{Meyers2006,Shiba2010,Yamakov2004,Shan2004,Cheng2009}.

A large number of numerical works have explored the microscopic dynamics induced in amorphous systems by a continuous shear, finding quite generally diffusive dynamics at the particle level~\cite{LemaitrePRL2009,TsamadosEPJE2010,BarratPRL2011}, once the affine component of the displacement is removed, as also confirmed by experiments on sheared colloidal glasses~\cite{PetekidisPRE2002,Besseling2007,Koumakis2012}. By contrast, the effect of a cyclic shear has been less investigated, in spite of its relevance to the fatigue tests commonly adopted in material science. Furthermore, cyclic deformation tests allow one to unambiguously identify irreversible rearrangements (as opposed to the non-affine displacement measured in continuous shear, which may be reversible) and to follow the evolution, or aging, of the dynamics as the sample is kept under an oscillatory deformation. Similarly to the case of continuous shear, the microscopic dynamics in cyclically deformed amorphous solids have been found to be diffusive (or even subdiffusive), in simulations~\cite{Fiocco2013,Priezjev2013} as well as in experiments on colloids~\cite{PetekidisPRE2002,ChenPRE2010} or granular matter~\cite{MartyPRL2005,SlotterbackPRE2012}. Aging effects have been reported for macroscopic quantities, \textit{e.g.} pressure or compacity~\cite{PouliquenPRL2003,RenPRL2013}, or for the microscopic dynamics. In the latter case, however, aging has been probed over a few tens of cycles at most~\cite{SlotterbackPRE2012,Keim2013}.

In this Letter, we report on experiments probing the irreversible rearrangements induced in a colloidal polycrystal by thousands of shear deformation cycles. Plasticity is investigated by confocal microscopy and by an ``echo'' light scattering technique, inspired by previous work on glassy colloids and emulsions~\cite{Hebraud1997,PetekidisPRE2002}. 
We find that plasticity slowly remodels the network of GBs via previously unreported ballistic dynamics. A steady state is eventually reached at all probed length scales, preceded by an aging regime whose duration is unexpectedly length-scale dependent.

\begin{figure}
\includegraphics[width=0.95\columnwidth,clip]{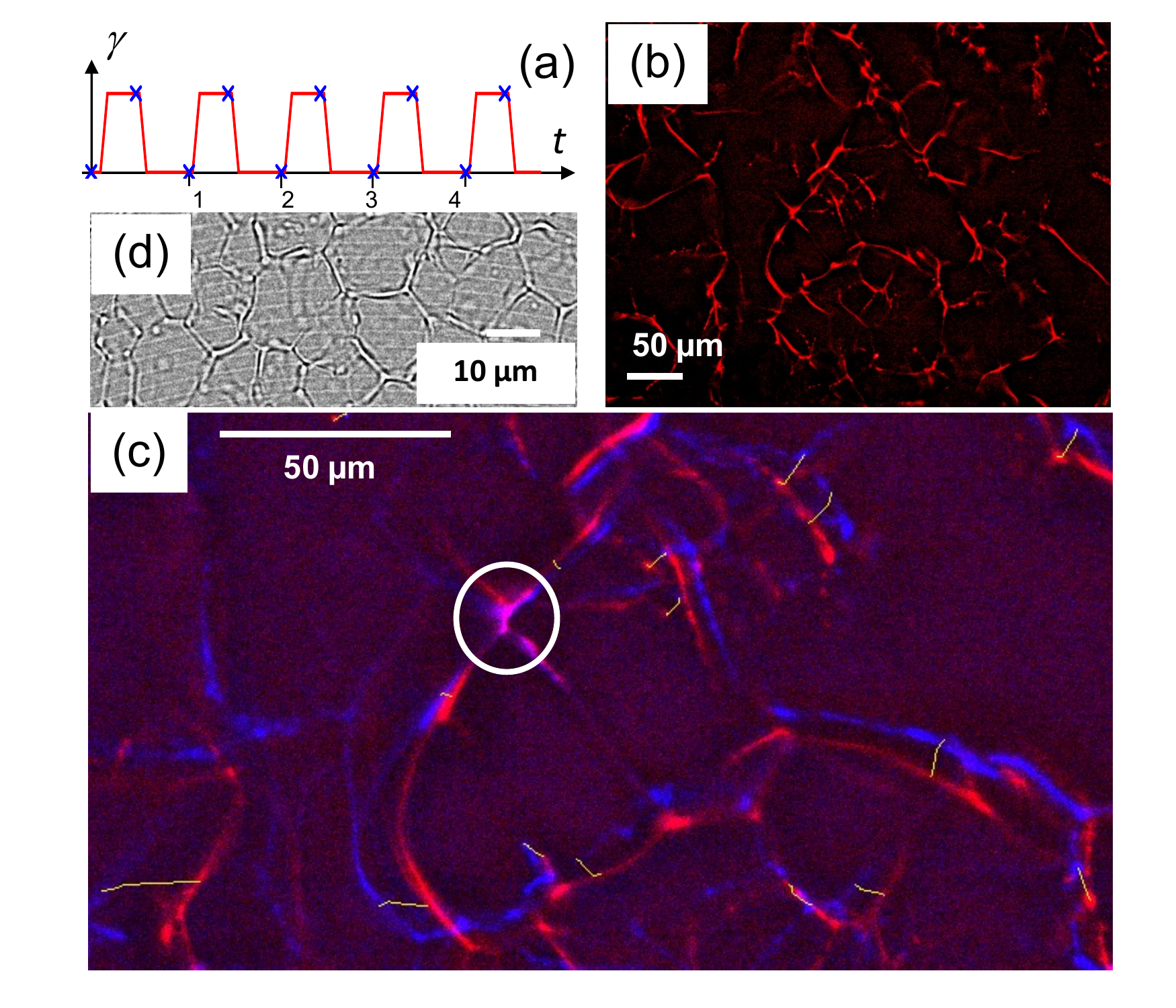}
\caption{(Color online) (a) Typical time-dependent shear deformation imposed to the sample. Time is expressed in units of the number of cycles, one cycle lasting 26 s. The crosses indicate when images are taken in microscopy or light scattering experiments. Confocal (b,c) and light (d) microscopy images of the grain boundary network of a colloidal polycrystal doped with fluorescent polystyrene (b,c) or silica (d) particles. (b) Image at rest; (c) Overlay of two images taken after $1$ (red) and $3711$ (blue) shear cycles of amplitude $\gamma = 3.6\%$. The yellow lines connect the position of representative GBs at times $t = 1, 112, 2617, 3130$, and 3711 cycles.}
\label{fig:fig1}
\end{figure}

The system~\cite{Tamborini2012a,Louhichi2013} comprises water-based thermosensitive block-copolymers (Pluronics F108) forming micelles arranged on a face-centered cubic lattice (lattice parameter $d_{\rm c}=32$ nm), to which a small amount of nanoparticles (NPs) of diameter $2a \approx d_{\rm c}$ is added. The NPs act as impurities, segregated in the GBs upon crystallization. The polycrystal is transparent to visible light; thus, microscopy and scattering experiments probe the network of GBs where the NP accumulate. The sample is prepared and injected in the cell in a fluid state at temperature $T \approx 2~^{\circ}\mathrm{C}$; crystallization is induced \textit{in situ }by rising $T$ up to $20~^{\circ}\mathrm{C}$. Thus, no uncontrolled shear is imposed when loading the sample, as it is often the case for pasty colloidal samples. The sample microstructure can be tuned by varying the NP concentration and the rate $\dot{T}$ at which the temperature is increased to induce crystallization~\cite{Ghofraniha2012}. Here, we fix $\dot{T} = 0.02 {^\circ}\mathrm{C}~\mathrm{min}^{-1}$. We impose a shear deformation by using a home-made shear cell~\cite{supplementary} where the sample is confined between two parallel glass plates separated by a gap $e$ ($e= 1.58~\mathrm{mm}$ and $250\um$ for light scattering and microscopy, respectively). A motor is used to displace one of the plates along the $x$ direction by an amount $\delta$, thereby imposing a strain $\gamma = \delta / e$. We cycle between sheared and unsheared states, as shown in Fig.~\ref{fig:fig1}(a). We express time in units of full cycles, whose duration is $26~\mathrm{s}$.

A confocal microscopy image of a sample doped with fluorescent polystyrene NPs ($2a=36~\mathrm{nm}$, volume fraction $\varphi = 0.05\%$) is shown in Fig.~\ref{fig:fig1}(b), where the network of GBs is clearly visible. 
For samples at rest, no evolution of the GB network is observed, even after several hours. This has to be contrasted with samples submitted to a cyclic shear. Figure \ref{fig:fig1}(c) shows an overlay of two images separated by a very large number (3710) of shear cycles.
The images overlap perfectly in regions where the deformation has been fully reversible, as in the zone highlighted by the white circle. However, in most of the field of view the images do not overlap, revealing GB migration, with displacements up to $\sim 10 ~\mu\mathrm{m}$, a sizeable fraction of the grain size. Both the magnitude and the direction of the GB displacement vary across the image, suggesting that plasticity involves a complex rearrangement of the whole network of GBs, and not just the sliding or rotation of the crystallites, as reported for metals~\cite{Margulies2001}. Remarkably, the GB trajectories are close to straight lines, ruling out shear-induced diffusive motion. The observation of plasticity at a microscopic level 
is consistent with oscillatory strain rheology measurements~\cite{supplementary} that probe macroscopically the mechanical response of the polycrystal, since in the range of $\gamma$ probed here the elastic limit is exceeded.

While confocal microscopy provides valuable insight on plasticity on the scale of a few grains, it cannot accurately measure small GB displacements when probing a large sample area~\cite{supplementary}. To provide a more quantitative account of the plasticity process, we couple the shear cell to a low-angle light scattering setup designed to access the characteristic length scale of the GB network~\cite{Tamborini2012b}. We check that the apparatus is stable enough to reliably probe the dynamics over thousands of cycles~\cite{supplementary}. We use a polycrystal doped with silica NPs ($2a = 30$ nm, $\varphi = 1\%$), yielding an average grain size of $10\um$ (Fig.~\ref{fig:fig1}(d)). A CCD camera records images of the speckle pattern scattered by the GB network at scattering vectors $q = 4\pi n \lambda^{-1} \sin (\theta/2)$ in the range $0.1\um^{-1} < q < 4\um^{-1}$, with $n=1.36$ the refractive index, $\lambda=632.8$ nm the in-vacuo laser wavelength and $\theta$ the scattering angle. The images are taken at each half cycle, while the sample is at rest (see Fig.~\ref{fig:fig1}(a)). Any rearrangement in the sample is mirrored by a change in the speckle images,  quantified by the two-time intensity correlation function:
\begin{equation}
g_2(t,\tau) - 1 =\frac{\langle I_p(t) I_p(t+\tau)\rangle _{\mathbf{q}}} {\langle I_p(t) \rangle _{\mathbf{q}} \langle I_p(t+\tau) \rangle _{\mathbf{q}}} - 1 = \beta f^2(\mathbf{q},t,\tau)\,.
\end{equation}
Here, $I_p(t)$ is the intensity of the $p-$th pixel at time $t$, $\langle \cdot \cdot \cdot \rangle _{\mathbf{q}}$ indicates an average over a set of pixels corresponding to a well-defined magnitude and orientation of $\mathbf{q}$, $\beta \lesssim 1$ is an instrumental constant~\cite{Berne1976}, and $f$ is the two-time intermediate scattering function that quantifies the particle displacement projected on $\mathbf{q}$~\cite{supplementary}. In the following, we show data for integer values of $\tau$, corresponding to unsheared states of the sample, and report results for the dynamics in the direction parallel to the imposed deformation ($\mathbf{q} // \mathbf{x}$); by analyzing DLS and microscopy~\cite{supplementary} data we have checked that the main findings are similar for $\mathbf{q} \bot \mathbf{x}$.

\begin{figure}
\includegraphics[width=0.95\columnwidth,clip]{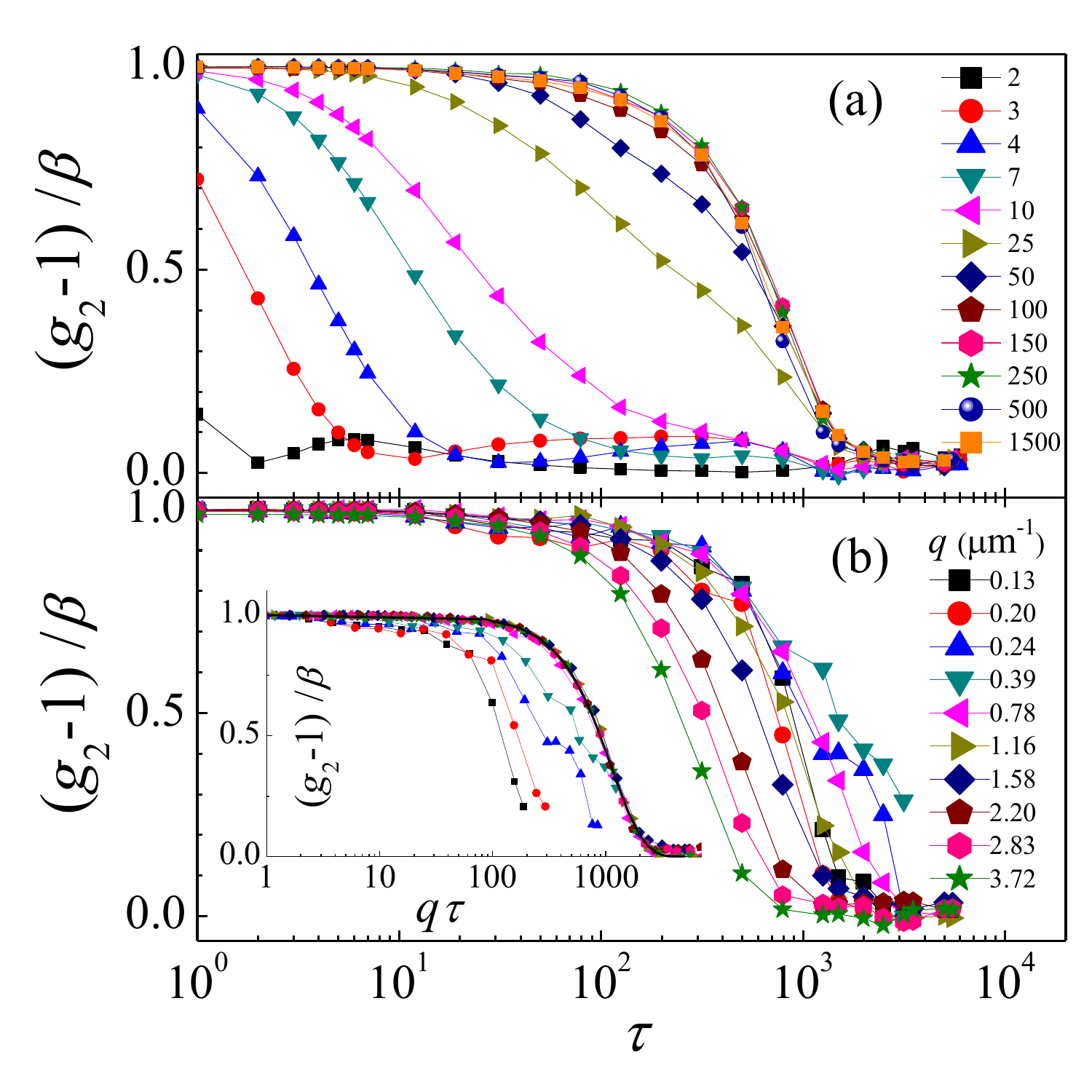}
\caption{(Color online) Intensity correlation functions for $\gamma = 4.6 \%$, (a) at a fixed wave vector $q=1.58\um^{-1}$ and for various aging times $t$, indicated by the legend, and (b) for various $q$'s, after a fixed number of cycles $t=500$, corresponding to the stationary regime. (Inset) Same data, plotted against $q \tau$. The continuous line is a compressed exponential fit yielding an exponent $p = 1.54$. }
\label{fig:fig2}
\end{figure}

Figure~\ref{fig:fig2} shows $\beta^{-1}(g_2-1)$ for a polycrystal submitted to shear cycles with strain amplitude $\gamma = 4.6\%$.  At short time lags, a high correlation level is measured, indicating that the GB network recovers its initial microscopic configuration after each shear cycle, hinting to a purely elastic behavior. Remarkably, however, the correlation functions always fully decay when probing the dynamics over a large number of cycles. This implies that the microscopic configuration of the GB network is eventually modified over distances  $\sim q^{-1}$ up to $8\um$, comparable to the grain size, an unambiguous signature of plasticity. We check by static light scattering and confocal microscopy that these rearrangements do not lead to a change of the average grain size, differently from Ref.~\cite{Gokhale2012}, where strain-induced grain growth was observed. Figure~\ref{fig:fig2}(a) shows data for a representative scattering vector and for various ages $t$, \textit{i.e.} after submitting the sample to $t$ deformation cycles. The shear-induced dynamics are non-stationary, since the decay time grows as $t$ increases.
However, a stationary state is eventually reached, since correlation functions for $t \geq 100$ overlap. We investigate the length scale dependence of the dynamics by plotting in Fig.~\ref{fig:fig2}(b) correlation functions measured simultaneously for various $q$-vectors in the stationary regime. Overall, the decay of $g_2-1$ shifts towards higher relaxation times when $q$ decreases, as expected because smaller $q$'s correspond to larger length scales. However, for $q \le q_\mathrm{c} \approx 0.5\um^{-1}$ the correlation functions depend only slightly on $q$, hinting at a peculiar length-scale dependence of the dynamics.

For $q > q_\mathrm{c}$, the correlation functions collapse onto a master curve when plotted \textit{vs} $q \tau$ (inset of Fig.~\ref{fig:fig2}(b)). This scaling rules out diffusive motion and unambiguously indicates that the GBs undergo ballistic displacements~\cite{Berne1976,supplementary}. Further insight on these very unusual dynamics can be obtained by analyzing the shape of $g_2-1$, which is very well fitted by a `compressed' exponential~\cite{Cipelletti2000,Cipelletti2003,bouchaud2001}, $g_2-1 = \exp[-(\tau/\tau_0)^p]$, with $p \approx 1.5 >1$, as opposed to the more usual $p<1$ in the stretched exponential relaxation of, \textit{e.g.}, glassy systems. The probability distribution function, $P_V(V_x)$,  of the $x$ component of the GB velocity can be obtained by Fourier transforming $f(q,\tau)$~\cite{supplementary}. We find that $P_V$ is a Levy law~\cite{bouchaud90}, a highly non-trivial distribution with a power law tail, $P_V(V_x) \sim |V_x|^{-(p+1)}$ at large $|V_x|$. The characteristic GB velocity is very small~\cite{supplementary}, of order $5.6\times10^{-4}\um$ per cycle, consistent with the fact that substantial GB motion occurs only after thousands of cycles. These results are further supported by the analysis of real space trajectories obtained by confocal microscopy~\cite{supplementary} and are also found for the other applied strains. Remarkably, the same $q\tau$ ballistic scaling and a similar compressed exponential shape have been reported for the spontaneous (non-driven) aging dynamics of a variety of out-of-equilibrium soft systems~\cite{Cipelletti2000,Cipelletti2003,ruzicka11,madsen10}, for which these peculiar features have been ascribed to the dipolar displacement field due to localized plastic rearrangements~\cite{bouchaud2001}. This analogy sheds light on the physical origin of the shear-induced dynamics in our colloidal polycrystal: the energy injected in the system by shearing does not act as a source of thermal-like noise, which would lead to diffusive motion. Rather, the dynamics result from plastic events, as in the predictions of~\cite{bouchaud2001}.  These events are likely to continuously trigger further events by redistributing stresses throughout the sample, a scenario envisioned by modern models of plasticity~\cite{baret2002,PicardPRE2005,homer2009}, thereby explaining why a stationary regime is eventually reached.

\begin{figure}
\includegraphics[width=0.95\columnwidth,clip]{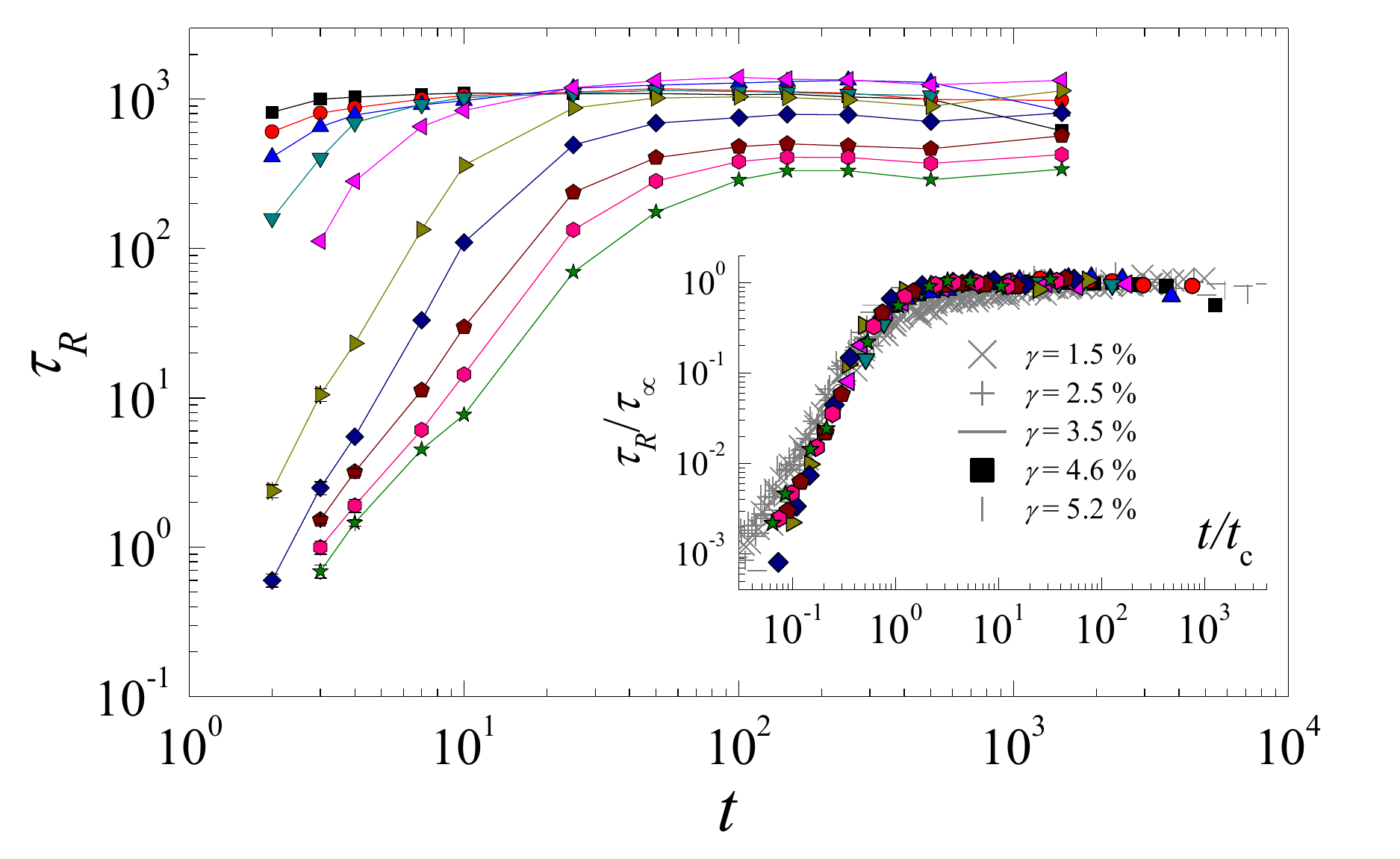}
\caption{(Color online) Evolution with the number of shear cycles of the characteristic relaxation time for various wave vectors $q$ (same symbols as in fig.~\ref{fig:fig2}b). The strain amplitude is $\gamma = 4.6 \%$. Inset: master curve obtained by using reduced variables, $\tau_R /\tau_{\infty}$ and $t/t_c$, for various strain amplitudes as indicated in the legend. The different solid symbols correspond to various $q$'s (same symbols as in the main plot).}
\label{fig:fig3}
\end{figure}

In order to quantify the dynamics beyond the stationary regime discussed so far, we determine the $q$-dependent characteristic relaxation time of the correlation functions, defined as  $\tau_R = \beta^{-1}\int \left[g_2(t,\tau)-1\right ]\mathrm{d}\tau$. We show in Fig.~\ref{fig:fig3} the evolution of $\tau_R$ with $t$ for various scattering vectors, at a fixed strain amplitude, $\gamma = 4.6 \%$. For all $q$'s, $\tau_R$ initially increases with $t$ and then reaches a plateau after a critical number of shear cycles, corresponding to the dynamical steady state where ballistic dynamics are observed. 
The overall shape of $\tau_R(t)$ appears to be similar regardless of $q$, suggesting that data for difference scattering vectors may be collapsed onto a master curve by choosing suitably renormalized variables. We test successfully such a scaling by plotting $\tau_R ^*\equiv \tau_R / \tauinf$ \textit{vs} $t^* \equiv t / \tc$ (inset of Fig.~\ref{fig:fig3}), where the scaling parameters $\tauinf$ and $\tc$ are the relaxation time in the asymptotic, stationary regime (simply related to the compressed exponential fitting parameter $\tau_0$ introduced above~\cite{supplementary}) and the crossover time between the aging and the stationary regime, respectively. Remarkably, we find that the same aging master curve holds to a very good approximation irrespective of the amplitude of the applied strain, for $1.5\% \leq \gamma \leq 5.2\%$ (inset Fig.~\ref{fig:fig3}). This aging master curve highlights the complex dynamics of the polycrystals, characterized by marked aging dynamics ($\tau_R^* \propto t^{* \nu}$ with $\nu = 2.2\pm0.3$), followed by a steady state.

\begin{figure}
\includegraphics[width=0.95\columnwidth,clip]{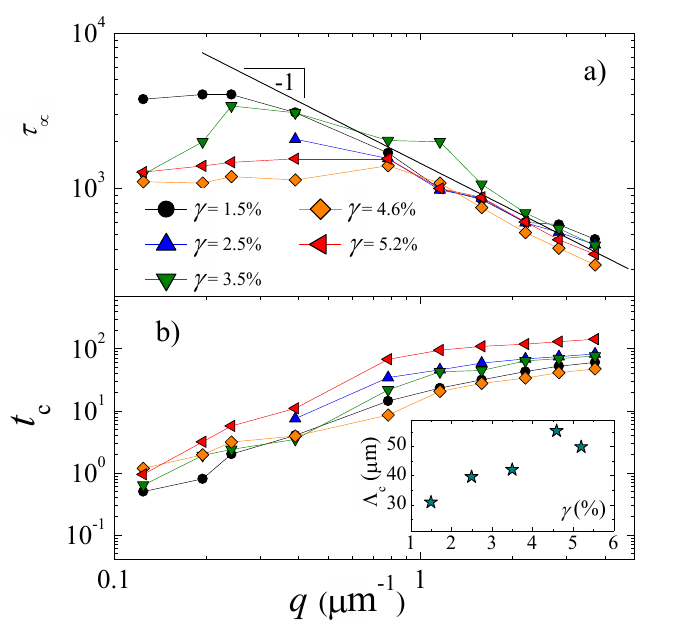}
\caption{(Color online) $q$ dependence of the normalization constants $\tau_{\infty}$ (a) and $t_c$ (b) used to obtain the master curve shown in Fig.~\ref{fig:fig3}. Data are labeled by the strain amplitude, as indicated in the legend. Inset of (b): $\gamma$ dependence of the critical length scale introduced in the text.}
\label{fig:fig4}
\end{figure}

The strain and $q$-dependence of the relaxation time in the steady state, $\tauinf$, is shown in Fig.~\ref{fig:fig4}(a). At large $q$, $\tauinf \sim q^{-m}$, with $m \approx 1$, a direct consequence of the $q \tau$ scaling discussed in reference to Fig.~\ref{fig:fig2}(b). In this regime, the relaxation time tends to decrease with increasing $\gamma$, a trend better seen when inspecting the characteristic GB velocity extracted from the corresponding velocity distributions (Table SM4 in~\cite{supplementary}). At low $q$, the same departure from the $\tauinf \sim q^{-1}$ scaling observed for $\gamma = 4.6\%$ in Fig.~\ref{fig:fig2} is also seen for all the other applied strains. Although the data are somehow scattered, they suggest that the characteristic relaxation time at low $q$ tends to increase as $\gamma$ decreases, while the crossover scattering vector appears to be rather $\gamma$-independent, with $2 \pi /q_{\rm c} \simeq 12\um$ on the order of the grain size ($10\um$).

The second scaling parameter, $\tc$, corresponds to the number of cycles needed to reach the steady state. Surprisingly, $\tc$ is found to steadily increase with $q$ (Fig.~\ref{fig:fig4}(b)), indicating that the time required to reach a steady state depends on the probed length scale, with stationary dynamics first attained on large length scales, a somehow counterintuitive result. To rationalize these findings, we assume that the applied shear allows the polycrystal to  explore regions in configuration space that were unaccessible to the spontaneous dynamics. Since the energy injected in the system by shearing is finite, these new configurations cannot be arbitrarily different from the initial ones; in particular, they must be closer to the initial ones at larger length scales, because the energetic cost of reconfiguring the sample over a length scale $\Lambda$ increases with $\Lambda$. Accordingly, stationary dynamics would be reached earlier at small $q$, because at large length scale the set of configurations explored in the stationary regime would be closer to the initial one. One can then define a critical length scale, $\Lambda_\mathrm{c}$, such that above it stationary dynamics are observed from the very beginning of the experiment (\textit{i.e.} from the first shear cycle). We take $\Lambda_c = 2 \pi / q^*$, where $q^*$ is the wave vector for which $\tc = 1$. The inset of Fig.~\ref{fig:fig4}(b) shows that $\Lambda_\mathrm{c}$ grows with the strain amplitude, consistent with the above picture, and is of the order of the grain size.

In conclusion, we have investigated plasticity in a cyclically sheared colloidal polycrystal. Our main finding is that shear-induced rearrangements are ballistic, a behavior at odd with previous simulations and with experiments on granular media and glassy colloids, for which diffusive dynamics under shear were reported. By contrast, both the ballistic dynamics and the compressed exponential relaxations found here are strongly reminiscent of the spontaneous dynamics of many out-of-equilibrium materials~\cite{Cipelletti2000,ruzicka11,Cipelletti2003,madsen10}, in agreement with mesoscopic models~\cite{bouchaud2001} and ongoing simulations~\cite{barrat2014} where the dynamics results from the dipolar strain field set by localized plastic events. Finally, the transition between the aging and the stationary regime exhibits an intriguing length scale dependence that, to our knowledge, has not been reported previously. 
More theoretical and experimental work will be needed to fully elucidate these surprising features.

\begin{acknowledgments}{We thank T. Phou and G. Pr\'{e}vot for help with instrumentation, M. George for the roughness measurements, S. Aime for help in microscopy experiments and L. Berthier and J.-L. Barrat for discussions. This work has been supported by ANR under Contract No. ANR-09-BLAN-0198 (COMET).}
\end{acknowledgments}


\pagebreak

In this Supplemental Material we briefly discuss:
\begin{itemize}
  \item  the shear cell design;
  \item  the measurements on purely elastic samples, to test the stability of the light scattering apparatus;
  \item  some rheological properties of the polycrystals (strain sweep);
  \item  how the distribution function of the grain boundary (GBs) velocity is retrieved from the dynamic structure factor measured by light scattering;
  \item  a comparison between light scattering and microscopy data, showing that both techniques lead to consistent results, both qualitatively (ballistic nature of the dynamics) and quantitatively (velocity distribution of the GBs)
\end{itemize}

\section{Shear cell design}

The shear cell is composed of two parallel microscope slides, which are sand-blasted (rms roughness $1\um$) to prevent slipping, except for a small window of diameter $\approx 2~\mathrm{mm}$ to probe optically the sample. For light scattering experiments, the spacing between the two slides is controlled by three stainless steel, high-precision ball bearings. The ball bearings are embedded in a custom-made rectangular frame of polydimethylsiloxane (PDMS), which contains the sample and avoids solvent evaporation. The PDMS frame is prepared by mixing two fluid components (a base and a cross-linker) with a mass-ratio 50:1, yielding a material with an elastic modulus of about 10 kPa. For microscopy observations, the two microscope slides are separated by a $250~\mu\mathrm{m}$ thick, $16 \times 16 \mathrm{mm}^2$ double-adhesive gene frame (Thermo Scientific), which acts as a spacer and avoids evaporation. A motor is used to displace one of the plates along the $x$ direction by an amount $\delta$, thereby imposing a strain $\gamma = \delta / e$, with $e$ the sample thickness. The motor speed during the displacement is $0.05~\mathrm{mm}~\mathrm{s}^{-1}$.

For both light scattering and microscopy, we measure the thickness of the sample chamber by microscopy, by measuring $e\mathrm{_{obj}}$, the vertical displacement of the microscope objective when focusing the upper and lower plates, respectively, using a $20\times$ air objective. The sample thickness is obtained as $e = n e\mathrm{_{obj}}$, where $n = 1.38$ is the sample refractive index. We measure $e\mathrm{_{obj}}$ at several locations separated by 1 cm, finding no difference to within the measurement uncertainty. This implies that the maximum deviation from parallelism over 1 cm is smaller than the measurement uncertainty ($\le 8~\mu\mathrm{m}$), corresponding to less than  $5\times 10^{-4}e$ (respectively, $1.3\times 10^{-3}e$) over the region sampled by light scattering (respectively, microscopy).

\section{Light scattering measurements on a purely elastic sample cyclically sheared}

We test our light scattering set-up  by measuring the correlation function $g_2 - 1$ for a sample whose mechanical response is purely elastic, a transparent PDMS elastomer, seeded with copper particles of diameter $3~\mu\mathrm{m}$. Here, the scattering signal is dominated by the contribution of the particles, whose microscopic configuration is essentially frozen due to the stiffness of the elastomer, whose elastic modulus is $G' \approx 500~\mathrm{kPa}$. We impose a cyclic shear deformation of amplitude $\gamma = 4.6$ \%. The inset of fig.~SM1 shows $\beta^{-1}(g_2-1)$ at short time lags, $\tau$: when $\tau$ corresponds to an odd number of half-cycles, the correlation drops to zero, due to the relative motion of the scatterers associated with the affine displacement field induced by the applied strain. For a delay time equal to an integer number of cycles, a high correlation level is recovered, an echo effect similar to that reported previously in concentrated emulsions and colloidal glasses~\cite{Hebraud1997,PetekidisPRE2002}. In the inset of Fig.~SM1 the height of the echos is unity, indicating that the scatterers have recovered exactly the initial microscopic configuration, as expected for tracer particles embedded in a purely elastic matrix, whose deformation is fully reversible. In the main figure, only data for integer values of $\tau$ are plotted, but for a very extended range of delay $\tau$. No significant loss of correlation is observed over $2000$ cycles, thus demonstrating that the setup is stable enough for the dynamics to be reliably probed over thousands of cycles.

\begin{figure}
\includegraphics[width=10cm]{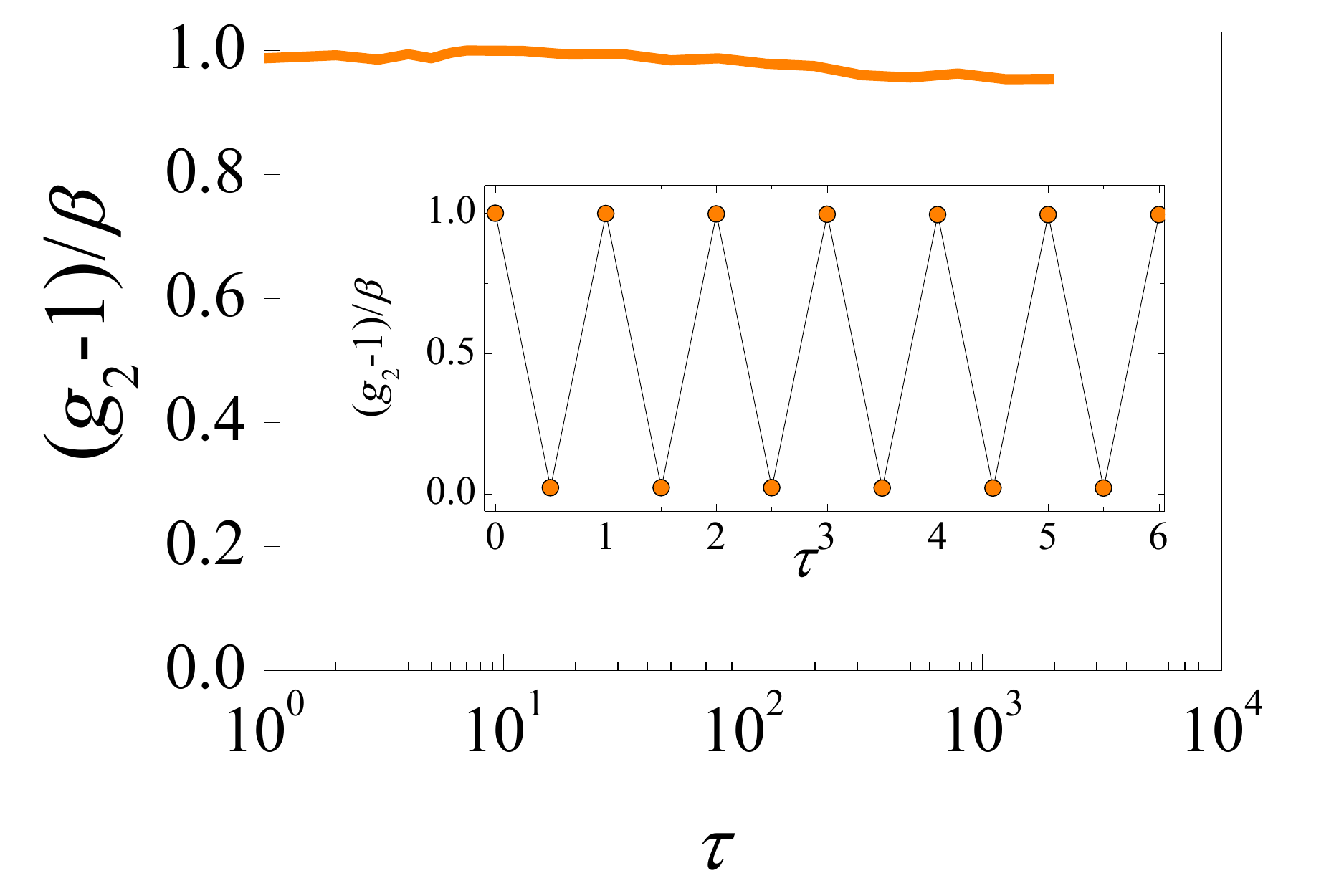}
\begin{center}
\noindent \textbf{Figure SM1:} (Color online) Intensity correlation functions for an elastic PDMS elastomer. Main figure: data for integer delays $\tau$.  Inset: zoom on the behavior of $\beta^{-1}(g_2-1)$ at small $\tau$. Data for both half-integer and integer delays are shown.
\label{fig:SM1}
\end{center}
\end{figure}

\section{Strain-dependence of the viscoelasticity of a colloidal polycrystal}

\begin{figure}[h!]
\includegraphics[width=10cm]{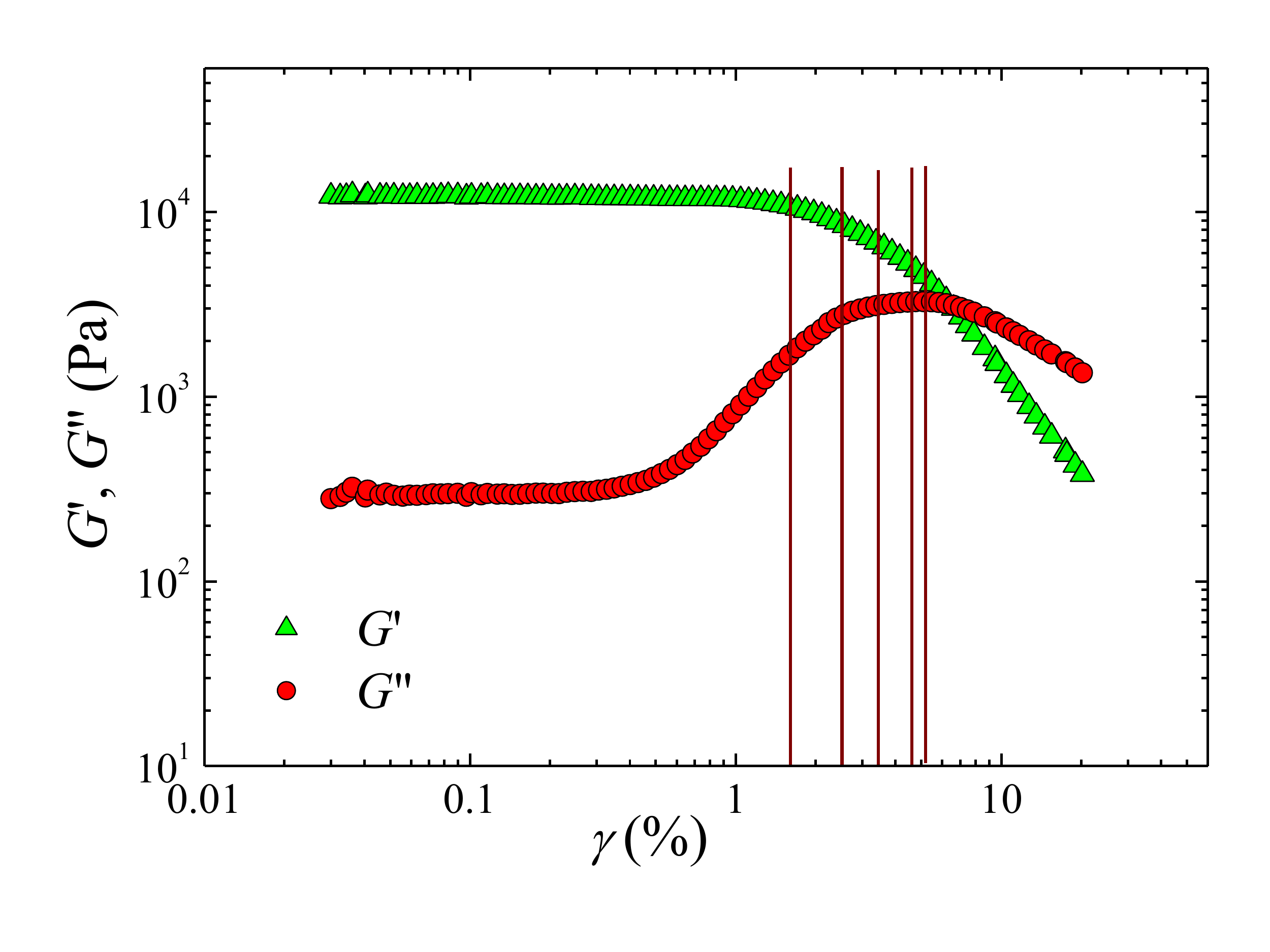}
\begin{center}
\noindent \textbf{Figure SM2:} (Color online) Strain dependence of the storage modulus, $G'$  (green triangles), and the loss modulus, $G''$ (red circles), as measured by standard oscillatory  rheology, at a fixed frequency, $f=0.5$ Hz. Same sample as for the light scattering experiments (micellar polycrystal doped with silica nanoparticles of diameter 30 nm, at a volume fraction $\varphi =1\%$). The vertical lines indicate the five strain amplitudes used in the light scattering experiments discussed in the main text.
\label{fig:SM2}
\end{center}
\end{figure}

Figure SM2 shows the strain dependence of the elastic and loss moduli measured by oscillatory rheology, for the colloidal polycrystal investigated by light scattering in the main text. The vertical lines indicate the strain amplitudes in the light scattering experiments: they correspond to an intermediate regime beyond the linear regime (which ends beyond $\gamma \approx 0.3\%$), but before fluidization occurs, for $\gamma \gtrsim 6 \%$.
\\

\section{Probability distribution function of the scatterers' velocity}

We show here that in the asymptotic, stationary regime the grain boundaries undergo ballistic motion and that the probability distribution function (PDF) of their velocity is a Levy stable law~\cite{bouchaud90}. In our dynamic light scattering (DLS) experiments we measure the intensity correlation function $g_2(q,\tau)-1$ (see Eq.~(1) of the main text), which is related to the intermediate scattering function $f(q,\tau)$ (also known as the dynamic structure factor) by $$g_2(\mathbf{q},\tau)-1 = \beta f(\mathbf{q},\tau)^2 \,,$$ with
\begin{equation}
\label{eqn:fq}
f(\mathbf{q},\tau) = \frac{1}{N}\left < \sum_{j,k=1}^N \exp \left \{i\mathbf{q}\cdot[\mathbf{r}_j(0)-\mathbf{r}_k(\tau)]\right\} \right > \,,
\end{equation}
with $N$ the number if scatterers, $\mathbf{r}_j$ the time-dependent position of the $j$-th scatterer, and where the brackets denote an ensemble average, and the factor $\beta$ an instrumental constant $\lesssim 1$~\cite{Berne1976}.

As discussed in the main text, we find that in the stationary regime correlation functions measured at different $q$ all collapse on a master curve when plotted against time scaled by $q \equiv |\mathbf{q}_x|$, for $q \ge q_\mathrm{c}$. This implies that $f(q,\tau)$ does not depend on $q$ and $\tau$ separately, but rather on the product $u = q \tau$. Recalling the compressed exponential shape of $g_2-1$, one has $f(q,\tau) = f(u) = \exp[-(uV_{0,x})^p]$, where $V_{0,x}$ represents the modulus of a characteristic velocity related to the relaxation time $\tau_R$ introduced in the main text by
\begin{equation}
\label{eq:V0}
V_{0,x} = \frac{\frac{2^{-\frac{1}{p}}}{p}\Gamma\left (\frac{1}{p} \right )}{q\tau_R}\,,
\end{equation}
with $\Gamma$ the gamma function.

\begin{figure}[h!]
\includegraphics[width=10cm]{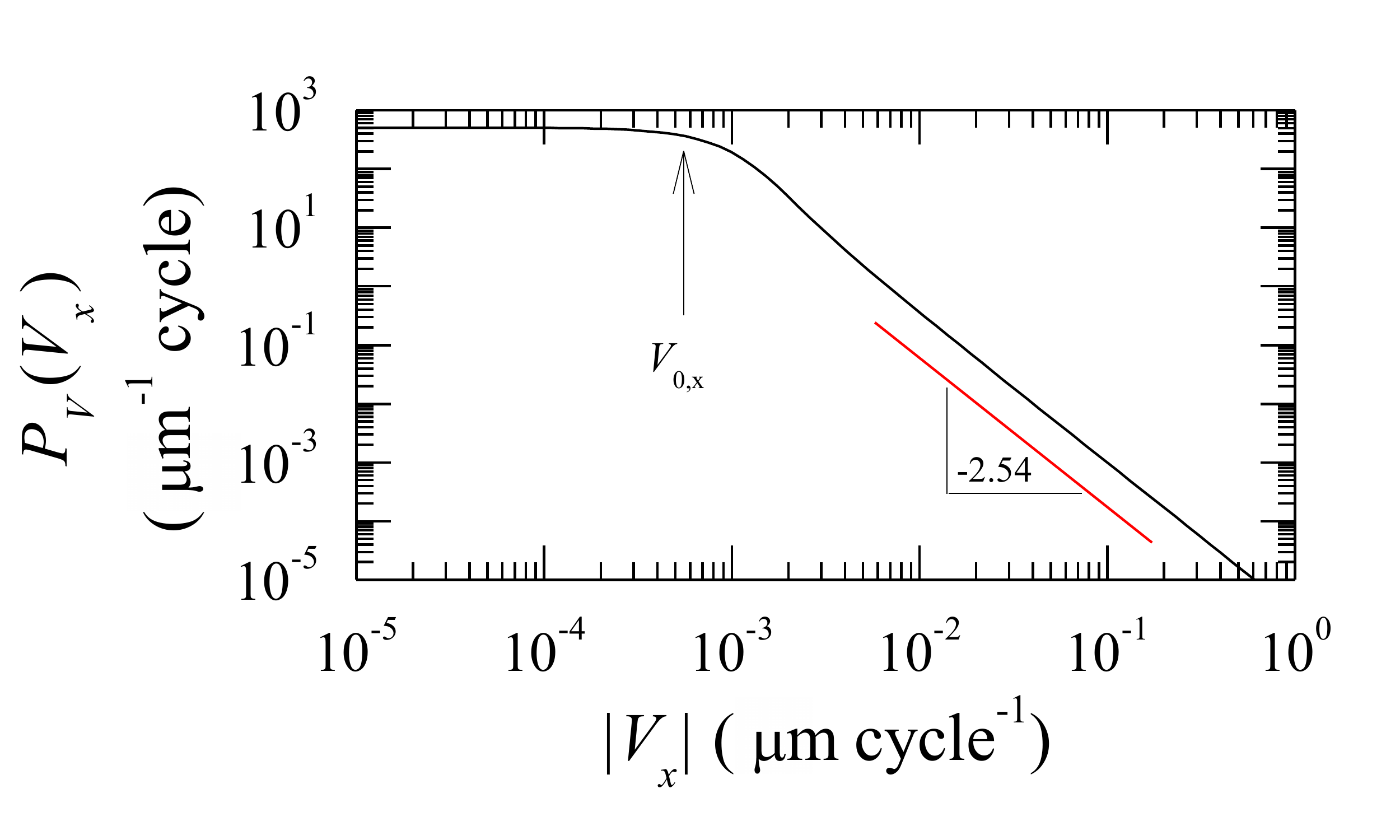}
\begin{center}
\noindent \textbf{Figure SM3:} (Color online) Probability distribution function of the $x$ component of the velocity of the grain boundaries in the asymptotic, stationary regime, for $\gamma = 4.6\%$. The PDF is obtained by Fourier transforming the compressed exponential fit to the data shown in the inset of Fig. 2b of the main text, see Eq.~(\ref{eq:pdf}) above. The distribution function is essentially flat up to the characteristic velocity $V_{0,x}$, while it decays as a power law for large $|V_x|$, with an exponent $-p-1$ directly related to the compressing exponent $p = 1.54$ of the fit to $g_2-1$. $V_{0,x}= 5.54~\mu\mathrm{m}~\mathrm{cycle}^{-1}$ is related to the $q$-dependent decay time of $g_2-1$, $\tau_R$, by Eq.~(\ref{eq:V0}).
\label{fig:SM3}
\end{center}
\end{figure}

Under these conditions, by following~\cite{Berne1976,LucaFaraday2003} one finds that the ensemble average in Eq.~(\ref{eqn:fq}) can be recast as an average over the probability distribution function of the $x$ component of the scatterers' velocity, $P_V(V_x)$, yielding:
\begin{equation}
\label{eq:fu}
f(u) = \int \mathbf{d}V_x P_V(V_x)\exp(-iuV_x) \,.
\end{equation}
By taking the inverse Fourier transform of Eq.~(\ref{eq:fu}), one finds
\begin{equation}
\label{eq:pdf}
P_V(V_x) \propto \int \mathbf{d}u f(u)\exp(iuV_x)= \int \mathbf{d}u \exp[-(uV_{0,x})^p] \exp(iuV_x)  \,.
\end{equation}
The last term on the r.h.s. of Eq.~(\ref{eq:pdf}) is the integral representation of the Levy stable law $L_{p,0}$~\cite{LucaFaraday2003,bouchaud90}. The Levy PDF is characterized by a flat distribution for $|V_x| \lesssim V_{0,x}$, followed by a power-law tail, $P_V(|V_x|) \sim |V_x|^{-p-1}$~\cite{bouchaud90}. We show in Fig. SM3 the PDF obtained by numerical integration of Eq.~(\ref{eq:pdf}), for the dynamics measured in the asymptotic regime shown in Fig. 2b of the main text ($\gamma = 4.6\%$, $p = 1.54$, $V_{0,x} = 5.54\times 10^{-4}~\mu\mathrm{m}~\mathrm{cycle}^{-1}$).

\begin{table} [h!]
\begin{center}
\begin{tabular}{l|l|l}
\multicolumn{1}{c|}{$\gamma$ (\%)} &
\multicolumn{1}{c|}{$p$} &
\multicolumn{1}{c}{$V_{0,x}~\mathrm{(}\mu\mathrm{m}~\mathrm{cycle}^{-1}\mathrm{)}$} \\
\hline
1.5 &	1.66 & $3.52 \times 10^{-4}$\\
2.5	& 1.76 & $2.82 \times 10^{-4}$\\
3.5	& 1.65 & $2.55 \times 10^{-4}$\\
4.6	& 1.54 & $5.64 \times 10^{-4}$\\
5.2	& 1.53 & $6.54 \times 10^{-4}$\\
\end{tabular}
\end{center}
\vspace{0.1 cm}
{\noindent \textbf {Table SM4:}
Compressing exponent $p$ governing the power-law tail of the velocity distribution of the grain boundaries and characteristic velocity obtained from the fits to the intensity correlation functions $g_2-1$ in the asymptotic, stationary regime, for the five values of the applied strain amplitude.}
\end{table}

Table SM4 summarizes the parameters of the Levy distributions of $V_x$ for all our experiments. We find that the $p$ exponent governing the slope of the tail of $P_V$ varies only slightly with the applied strain, $\gamma$. The variation of the characteristic velocity is more pronounced: the general trend is for $V_{0,x}$ to increase with $\gamma$, albeit with some scatter in the data. The order of magnitude of $V_{0,x}$ is $5 \times 10^{-4}\mu\mathrm{m}~\mathrm{cycle}^{-1}$. This is close to $V_{0,x} \sim 10^{-3}\mu\mathrm{m}~\mathrm{cycle}^{-1}$, as obtained by analyzing in real space the grain boundary displacement for a sample with a slightly larger grain size, as discussed in Sec.~\ref{sec:cm_vs_dls} below.

\section{Comparison between light scattering and microscopy measurements}
\label{sec:cm_vs_dls}

\begin{figure}
\includegraphics[width=12.5cm]{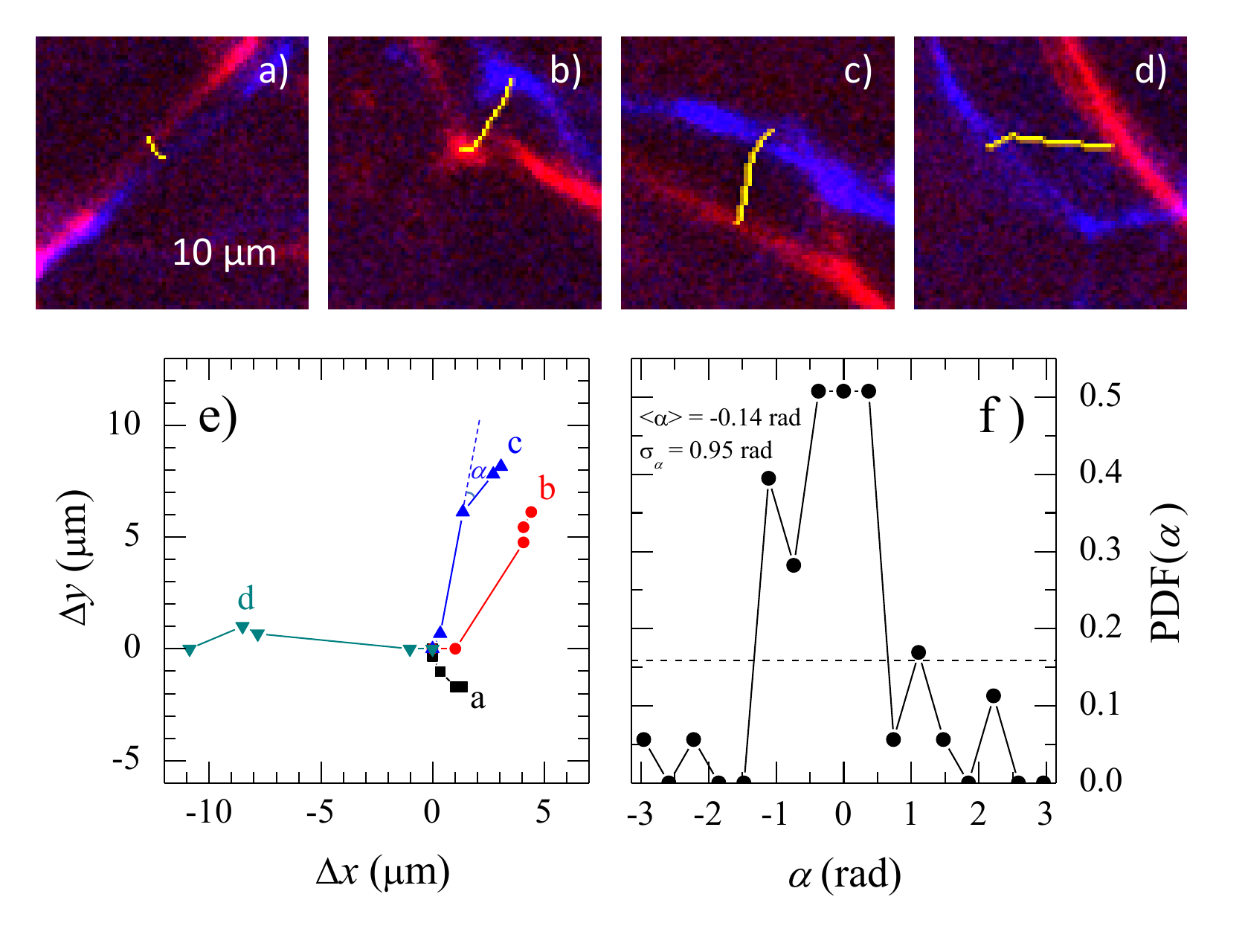}
\begin{center}
\noindent \textbf{Figure SM5} (Color online) a) - d): zoom into representative GB trajectories, from Fig. 1c of the main text. The trajectories are obtained by measuring the GB position at $t = 1, 112, 2617, 3130$, and 3711 shear cycles. e): GB displacement with respect to the position at $t=1$, for the trajectories shown in a)-d), as indicated by the labels. The angle $\alpha$ between consecutive segments of a trajectory is also shown for two segments of the trajectory c). f): probability distribution function of $\alpha$. The PDF is strongly peaked around $\alpha = 0$, implying that the GBs trajectories are close to straight lines, a behavior suggestive of ballistic motion and incompatible with diffusion, for which $\alpha$ would be evenly distributed (dotted line). The labels indicate the average and the standard deviation of $\alpha$.
\label{fig:SM5}
\end{center}
\end{figure}

In order to provide additional support to the analysis of the grain boundary motion performed on light scattering data, we measure the GB displacement also in microscopy experiments. Figures SM5a-d zoom into some of the trajectories shown in Fig. 1c of the main text. The trajectories are obtained by measuring the position of representative GBs at times $t = 1, 112, 2617, 3130$, and 3711 cycles. The trajectories are overlaid to the images of the polycrystal taken at $t=1$ (red) and $3711$ (blue). The images at intermediate times are not shown for clarity. For the same GBs, the displacement $(\Delta x,\Delta y)$ with respect to the position at $t=1$ is shown in Fig. SM5e. Clearly, the trajectories are close to straight lines, a behavior incompatible with random motion. To quantify the tendency of the GBs to move along straight lines, we calculate the angle $\alpha$ between successive segments of the trajectories, as exemplified in Fig. SM5e. Figure SM5f shows the PDF of $\alpha$, obtained from all the trajectory segments at all the locations shown in Fig. 1c of the main text, \textit{i.e.} 52 segments from 13 different trajectories. The PDF is strongly peaked around $\alpha = 0$, confirming that the trajectories are incompatible with diffusive motion and are rather suggestive of straight-line displacements as in ballistic motion, consistent with the DLS results.

\begin{figure}[h!]
\includegraphics[width=9cm]{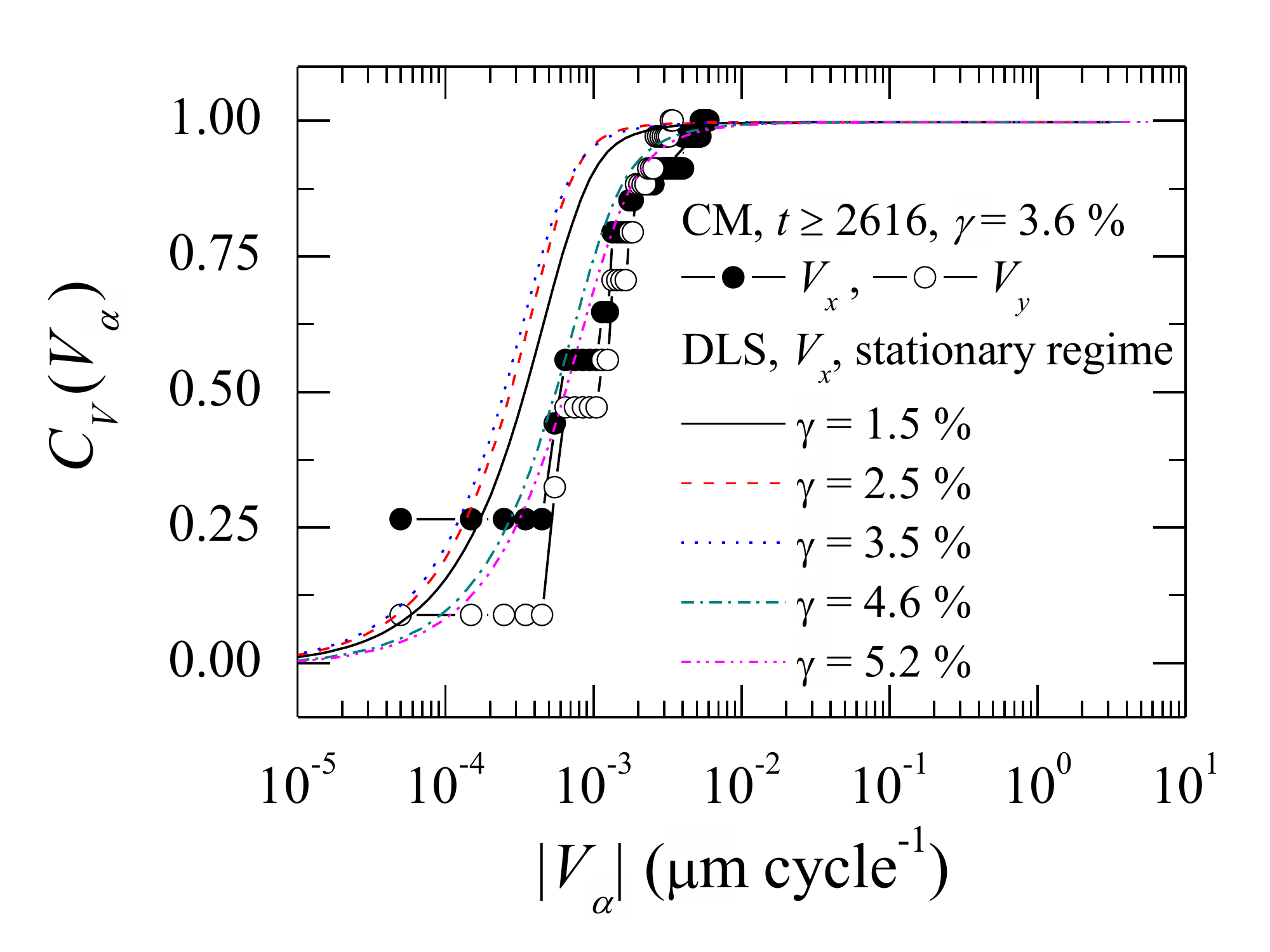}
\begin{center}
\noindent \textbf{Figure SM6:} (Color online) Cumulative distribution function of the velocity components $V_\alpha$ of the grain boundaries in the asymptotic, stationary regime ($\alpha = x$ (resp. , $y$) for the component parallel (resp., perpendicular) to the direction of the applied strain). CM and symbols: data obtained by analyzing the trajectories obtained from confocal microscopy and shown in Fig. 1c of the main text ($\gamma = 3.6\%$). Lines: data obtained by light scattering for the five strain amplitudes reported in the main text.
\label{fig:SM3}
\end{center}
\end{figure}

We measure the GB velocity over two time intervals, $t\in [2617-3130]$ and $t\in [3130-3711]$. We find comparable average velocities, consistently with the notion that the sample dynamics become stationary at large $t$, as seen by DLS.
We calculate the cumulative distribution of the GB velocity, using all trajectories and both time intervals and compare it to the cumulative velocity distributions obtained from the DLS data analysis. The results are shown in Fig. SM6 for the components of the velocity parallel and perpendicular to the shear direction.

Several comments are in order. First, there is an overall good agreement between the microscopy and DLS data. This agreement is particularly remarkable given that the sample composition (nanoparticle kind, size and concentration) has been separately optimized according to the specific requirements of each experiment. As a consequence, the grain size --although of the same order of magnitude-- is not identical in the samples used for microscopy and light scattering (see Figs. 1b and 1d in the main text). Second, microscopy data measured for the $x$ and $y$ direction overlap, thus indicating that plasticity is essentially isotropic. Third, the order of magnitude of the GB velocity in the stationary regime is very small, as also seen in Figs. SM3 and SM5. This highlights a key requirement of our experiments, \textit{i.e.} sensitivity to small-scale motion. In this respect, light scattering is superior to microscopy: for the former, the smallest rms displacement that can be reliably measured is $\Delta r_{\mathrm{min}} \sim 0.1\um$, corresponding to a decay of 5\% of $g_2-1$ at the largest scattering vector. For microscopy, $\Delta r_{\mathrm{min}} \sim 0.34 \um$ (corresponding to 1 pixel), more than three times larger than by DLS. An additional advantage of light scattering is a better statistics: for DLS, the probed volume is $V_{\mathrm{scatt}} \sim 1 ~\mathrm{mm}^3$, about 80 times larger than $V_{\mathrm{micro}} \sim 0.013~\mathrm{mm}^3$, the volume accessible to confocal microscopy. These advantages motivate our choice of DLS as the main quantitative probe of the GB dynamics.


\begin{thebibliography}{33}
\expandafter\ifx\csname natexlab\endcsname\relax\def\natexlab#1{#1}\fi
\expandafter\ifx\csname bibnamefont\endcsname\relax
  \def\bibnamefont#1{#1}\fi
\expandafter\ifx\csname bibfnamefont\endcsname\relax
  \def\bibfnamefont#1{#1}\fi
\expandafter\ifx\csname citenamefont\endcsname\relax
  \def\citenamefont#1{#1}\fi
\expandafter\ifx\csname url\endcsname\relax
  \def\url#1{\texttt{#1}}\fi
\expandafter\ifx\csname urlprefix\endcsname\relax\def\urlprefix{URL }\fi
\providecommand{\bibinfo}[2]{#2}
\providecommand{\eprint}[2][]{\url{#2}}




\bibitem{LemaitrePRL2009} A. Lema\^{\i}tre and C. Caroli, Phys. Rev. Lett.,  \textbf{103}, 065501, 2009.


\bibitem{TsamadosEPJE2010} M. Tsamados, Eur. Phys. J. E,  \textbf{32}, 165, 2010.

\bibitem{BarratPRL2011} K. Martens, L. Bocquet and J.-L. Barrat, Phys. Rev. Lett.,  \textbf{106}, 156001, 2011.

\bibitem[{\citenamefont{Falk}(1998)\citenamefont{Falk}}]{Falk1998}
\bibinfo{author}{\bibfnamefont{M. L.}~\bibnamefont{Falk}},
\bibnamefont{and} \bibinfo{author}{\bibfnamefont{J. S.} \bibnamefont{Langer}},
\bibinfo{journal}{Phys. Rev. E} \textbf{\bibinfo{volume}{57}},
\bibinfo{pages}{7192} (\bibinfo{year}{1998}).

\bibitem[{\citenamefont{Bocquet}(2009)\citenamefont{Bocquet}}]{Bocquet2009}
\bibinfo{author}{\bibfnamefont{L.}~\bibnamefont{Bocquet}},
\bibinfo{author}{\bibfnamefont{A.}~\bibnamefont{Colin}},
\bibnamefont{and} \bibinfo{author}{\bibfnamefont{A.} \bibnamefont{Ajdari}},
\bibinfo{journal}{Phys. Rev. Lett.} \textbf{\bibinfo{volume}{103}},
\bibinfo{pages}{036001} (\bibinfo{year}{2009}).

\bibitem{Jop2012} P. Jop, V. Mansard, P. Chaudhuri, L. Bocquet, and A. Colin, Phys. Rev. Lett.,  \textbf{108}, 148301, 2012.

\bibitem{Koumakis2012} N. Koumakis, M. Laurati, S. U. Egelhaaf, J. F. Brady, and G. Petekidis, Phys. Rev. Lett.,  \textbf{108}, 098303, 2012.

\bibitem[{\citenamefont{Biswas}(2013)\citenamefont{Biswas}}]{Biswas2013}
\bibinfo{author}{\bibfnamefont{S.}~\bibnamefont{Biswas}},
\bibinfo{author}{\bibfnamefont{M.}~\bibnamefont{Grant}},
\bibinfo{author}{\bibfnamefont{I.}~\bibnamefont{Samajdar}},
\bibinfo{author}{\bibfnamefont{A.}~\bibnamefont{Haldar}},
\bibnamefont{and} \bibinfo{author}{\bibfnamefont{A.} \bibnamefont{Sain}},
\bibinfo{journal}{Scientific Reports} \textbf{\bibinfo{volume}{3}},
\bibinfo{pages}{2728} (\bibinfo{year}{2013}).

\bibitem{Zhang2009} H.Zhang, D. J. Srolovitz, J. F. Douglas, and J. A. Warren, Proc. National Acad. Sci. USA,  \textbf{106}, 7735, 2009.

\bibitem{Nagamanasa2011} K. H. Nagamanasa, S. Gokhale, R. Ganapathy, and A. K. Sood, Proc. National Acad. Sci. USA,  \textbf{108}, 11323, 2011.

\bibitem[{\citenamefont{Shiba}(2010)\citenamefont{Shiba}}]{Shiba2010}
\bibinfo{author}{\bibfnamefont{H.}~\bibnamefont{Shiba}},
\bibnamefont{and} \bibinfo{author}{\bibfnamefont{A.} \bibnamefont{Onuki}},
\bibinfo{journal}{Phys. Rev. E} \textbf{\bibinfo{volume}{81}},
\bibinfo{pages}{051501} (\bibinfo{year}{2010}).

\bibitem[{\citenamefont{Meyers}(2006)\citenamefont{Meyers}}]{Meyers2006}
\bibinfo{author}{\bibfnamefont{M. A.}~\bibnamefont{Meyers}},
\bibinfo{author}{\bibfnamefont{A.}~\bibnamefont{Mishra}},
\bibnamefont{and} \bibinfo{author}{\bibfnamefont{D. J.} \bibnamefont{Benson}},
\bibinfo{journal}{Progress in Materials Science} \textbf{\bibinfo{volume}{51}},
\bibinfo{pages}{427} (\bibinfo{year}{2006}).

\bibitem[{\citenamefont{Yamakov}(2004)\citenamefont{Yamakov}}]{Yamakov2004}
\bibinfo{author}{\bibfnamefont{V.}~\bibnamefont{Yamakov}},
\bibinfo{author}{\bibfnamefont{D.}~\bibnamefont{Wolf}},
\bibinfo{author}{\bibfnamefont{S. R.}~\bibnamefont{Phillpot}},
\bibinfo{author}{\bibfnamefont{A. K.}~\bibnamefont{Mukherjee}},
\bibnamefont{and} \bibinfo{author}{\bibfnamefont{H.} \bibnamefont{Gleiter}},
\bibinfo{journal}{Nature Mater.} \textbf{\bibinfo{volume}{3}},
\bibinfo{pages}{43} (\bibinfo{year}{2004}).

\bibitem[{\citenamefont{Shan}(2004)\citenamefont{Shan}}]{Shan2004}
\bibinfo{author}{\bibfnamefont{Z.}~\bibnamefont{Shan}},
\bibinfo{author}{\bibfnamefont{E. A.}~\bibnamefont{Stach}},
\bibinfo{author}{\bibfnamefont{J. M. K.}~\bibnamefont{Wiezorek}},
\bibinfo{author}{\bibfnamefont{J. A.}~\bibnamefont{Knapp}},
\bibinfo{author}{\bibfnamefont{D. M.}~\bibnamefont{Follstaedt}},
\bibnamefont{and} \bibinfo{author}{\bibfnamefont{S. X.} \bibnamefont{Mao}},
\bibinfo{journal}{Science} \textbf{\bibinfo{volume}{305}},
\bibinfo{pages}{654} (\bibinfo{year}{2004}).

\bibitem[{\citenamefont{Cheng}(2009)\citenamefont{Cheng}}]{Cheng2009}
\bibinfo{author}{\bibfnamefont{S.}~\bibnamefont{Cheng}},
\bibinfo{author}{\bibfnamefont{A. D.}~\bibnamefont{Stoica}},
\bibinfo{author}{\bibfnamefont{X.-L.}~\bibnamefont{Wang}},
\bibinfo{author}{\bibfnamefont{Y.}~\bibnamefont{Ren}},
\bibinfo{author}{\bibfnamefont{J.}~\bibnamefont{Almer}},
\bibinfo{author}{\bibfnamefont{J. A.}~\bibnamefont{Horton}},
\bibinfo{author}{\bibfnamefont{C. T.}~\bibnamefont{Liu}},
\bibinfo{author}{\bibfnamefont{B.}~\bibnamefont{Clausen}},
\bibinfo{author}{\bibfnamefont{D. W.}~\bibnamefont{Brown}},
\bibinfo{author}{\bibfnamefont{P. K.}~\bibnamefont{Liaw}},
\bibnamefont{and} \bibinfo{author}{\bibfnamefont{L.} \bibnamefont{Zuo}},
\bibinfo{journal}{Phys. Rev. Lett.} \textbf{\bibinfo{volume}{103}},
\bibinfo{pages}{035502} (\bibinfo{year}{2009}).

\bibitem[{\citenamefont{Besseling}(2007)\citenamefont{Besseling}}]{Besseling2007}
\bibinfo{author}{\bibfnamefont{R.}~\bibnamefont{Besseling}},
\bibinfo{author}{\bibfnamefont{E. R.}~\bibnamefont{Weeks}},
\bibinfo{author}{\bibfnamefont{A. B.}~\bibnamefont{Schofield}},
\bibnamefont{and} \bibinfo{author}{\bibfnamefont{W. C. K.} \bibnamefont{Poon}},
\bibinfo{journal}{Phys. Rev. Lett.} \textbf{\bibinfo{volume}{99}},
\bibinfo{pages}{028301} (\bibinfo{year}{2007}).

\bibitem{PetekidisPRE2002} G. Petekidis, A. Moussaid and P. N. Pusey, Phys. Rev. E,  \textbf{66}, 051402, 2002.

\bibitem[{\citenamefont{Fiocco}(2013)\citenamefont{Fiocco}}]{Fiocco2013}
\bibinfo{author}{\bibfnamefont{D.}~\bibnamefont{Fiocco}},
\bibinfo{author}{\bibfnamefont{G.}~\bibnamefont{Foffi}},
\bibnamefont{and} \bibinfo{author}{\bibfnamefont{S.} \bibnamefont{Sastry}},
\bibinfo{journal}{Phys. Rev. E} \textbf{\bibinfo{volume}{88}},
\bibinfo{pages}{020301(R)} (\bibinfo{year}{2013}).

\bibitem{Priezjev2013} N. V. Priezjev, Phys. Rev. E,  \textbf{87}, 052302, 2013.

\bibitem{ChenPRE2010} D. Chen, D. Semwogerere, J. Sato, V. Breedveld and E. R. Weeks, Phys. Rev. E,  \textbf{81}, 011403, 2010.

\bibitem{MartyPRL2005} G. Marty and O. Dauchot, Phys. Rev. Lett.,  \textbf{94}, 2005.

\bibitem{SlotterbackPRE2012} S. Slotterback, M. Mailman, K. Ronaszegi, M. van Hecke, M. Girvan and W. Losert, Phys. Rev. E,  \textbf{85}, 021309, 2012.

\bibitem{PouliquenPRL2003} O. Pouliquen, M. Belzons and M. Nicolas, Phys. Rev. Lett.,  \textbf{91}, 2003.

\bibitem{RenPRL2013} J. Ren, J. A. Dijksman and R. P. Behringer, Phys. Rev. Lett.,  \textbf{110}, 018302, 2013.
\bibitem[{\citenamefont{Keim}(2013)\citenamefont{Keim}}]{Keim2013}
\bibinfo{author}{\bibfnamefont{N. C.}~\bibnamefont{Keim}},
\bibnamefont{and} \bibinfo{author}{\bibfnamefont{P. E.} \bibnamefont{Arratia}},
\bibinfo{journal}{Soft Matter} \textbf{\bibinfo{volume}{9}},
\bibinfo{pages}{6222} (\bibinfo{year}{2013}).

\bibitem[{\citenamefont{Hebraud}(1997)\citenamefont{Hebraud}}]{Hebraud1997}
\bibinfo{author}{\bibfnamefont{P.}~\bibnamefont{H\'{e}braud}},
\bibinfo{author}{\bibfnamefont{F.}~\bibnamefont{Lequeux}},
\bibinfo{author}{\bibfnamefont{J. P.}~\bibnamefont{Munch}},
\bibnamefont{and} \bibinfo{author}{\bibfnamefont{D. J.} \bibnamefont{Pine}},
\bibinfo{journal}{Phys. Rev. Lett.} \textbf{\bibinfo{volume}{78}},
\bibinfo{pages}{4657} (\bibinfo{year}{1997}).

\bibitem[{\citenamefont{Tamborini}(2012)\citenamefont{Tamborini}}]{Tamborini2012a}
\bibinfo{author}{\bibfnamefont{E.}~\bibnamefont{Tamborini}},
\bibinfo{author}{\bibfnamefont{N.}~\bibnamefont{Ghofraniha}},
\bibinfo{author}{\bibfnamefont{J.}~\bibnamefont{Oberdisse}},
\bibinfo{author}{\bibfnamefont{L.}~\bibnamefont{Cipelletti}},
\bibnamefont{and} \bibinfo{author}{\bibfnamefont{L.} \bibnamefont{Ramos}},
\bibinfo{journal}{Langmuir} \textbf{\bibinfo{volume}{28}},
\bibinfo{pages}{8562} (\bibinfo{year}{2012}).

\bibitem[{\citenamefont{Louhichi}(2013)\citenamefont{Louhichi}}]{Louhichi2013}
\bibinfo{author}{\bibfnamefont{A.}~\bibnamefont{Louhichi}},
\bibinfo{author}{\bibfnamefont{E.}~\bibnamefont{Tamborini}},
\bibinfo{author}{\bibfnamefont{N.}~\bibnamefont{Ghofraniha}},
\bibinfo{author}{\bibfnamefont{F.}~\bibnamefont{Caton}},
\bibinfo{author}{\bibfnamefont{D.}~\bibnamefont{Roux}},
\bibinfo{author}{\bibfnamefont{J.}~\bibnamefont{Oberdisse}},
\bibinfo{author}{\bibfnamefont{L.}~\bibnamefont{Cipelletti}},
\bibnamefont{and} \bibinfo{author}{\bibfnamefont{L.} \bibnamefont{Ramos}},
\bibinfo{journal}{Phys. Rev. E} \textbf{\bibinfo{volume}{87}},
\bibinfo{pages}{032306} (\bibinfo{year}{2013}).

\bibitem[{\citenamefont{Ghofraniha}(2012)\citenamefont{Ghofraniha}}]{Ghofraniha2012}
\bibinfo{author}{\bibfnamefont{N.}~\bibnamefont{Ghofraniha}},
\bibinfo{author}{\bibfnamefont{E.}~\bibnamefont{Tamborini}},
\bibinfo{author}{\bibfnamefont{J.}~\bibnamefont{Oberdisse}},
\bibinfo{author}{\bibfnamefont{L.}~\bibnamefont{Cipelletti}},
\bibnamefont{and} \bibinfo{author}{\bibfnamefont{L.} \bibnamefont{Ramos}},
\bibinfo{journal}{Soft Matter} \textbf{\bibinfo{volume}{8}},
\bibinfo{pages}{6214} (\bibinfo{year}{2012}).

\bibitem{supplementary} Supplemental Material at [link to be inserted by APS].

\bibitem[{\citenamefont{Margulies}(2001)\citenamefont{Margulies}}]{Margulies2001}
\bibinfo{author}{\bibfnamefont{L.}~\bibnamefont{Margulies}},
\bibinfo{author}{\bibfnamefont{G.}~\bibnamefont{Winther}},
\bibnamefont{and} \bibinfo{author}{\bibfnamefont{H. F.} \bibnamefont{Poulsen}},
\bibinfo{journal}{Science} \textbf{\bibinfo{volume}{291}},
\bibinfo{pages}{2392} (\bibinfo{year}{2001}).

\bibitem[{\citenamefont{Tamborini}(2012)\citenamefont{Tamborini}}]{Tamborini2012b}
\bibinfo{author}{\bibfnamefont{E.}~\bibnamefont{Tamborini}},
\bibnamefont{and} \bibinfo{author}{\bibfnamefont{L.}~\bibnamefont{Cipelletti}},
\bibinfo{journal}{Rev. Sci. Inst.} \textbf{\bibinfo{volume}{83}},
\bibinfo{pages}{093106} (\bibinfo{year}{2012}).

\bibitem{Berne1976} B. J. Berne and R. Pecora, \textit{Dynamic Light Scattering}, Wiley, New York, 1976

\bibitem[{\citenamefont{Gokhale}(2012)\citenamefont{Gokhale}}]{Gokhale2012}
\bibinfo{author}{\bibfnamefont{S.}~\bibnamefont{Gokhale}},
\bibinfo{author}{\bibfnamefont{K. H.}~\bibnamefont{Nagamanasa}},
\bibinfo{author}{\bibfnamefont{V.}~\bibnamefont{Santhosh}},
\bibinfo{author}{\bibfnamefont{A. K.}~\bibnamefont{Sood}},
\bibnamefont{and} \bibinfo{author}{\bibfnamefont{R.} \bibnamefont{Ganapathy}},
\bibinfo{journal}{Proc. National Acad. Sci. USA} \textbf{\bibinfo{volume}{109}},
\bibinfo{pages}{20314} (\bibinfo{year}{2012}).

\bibitem[{\citenamefont{Cipelletti}(2000)\citenamefont{Cipelletti}}]{Cipelletti2000}
\bibinfo{author}{\bibfnamefont{L.}~\bibnamefont{Cipelletti}},
\bibinfo{author}{\bibfnamefont{S.}~\bibnamefont{Manley}},
\bibinfo{author}{\bibfnamefont{R. C.}~\bibnamefont{Ball}},
\bibnamefont{and} \bibinfo{author}{\bibfnamefont{D. A.} \bibnamefont{Weitz}},
\bibinfo{journal}{Phys. Rev. Lett.} \textbf{\bibinfo{volume}{84}},
\bibinfo{pages}{2275} (\bibinfo{year}{2000}).

\bibitem[{\citenamefont{Cipelletti}(2003)\citenamefont{Cipelletti}}]{Cipelletti2003}
\bibinfo{author}{\bibfnamefont{L.}~\bibnamefont{Cipelletti}},
\bibinfo{author}{\bibfnamefont{L.}~\bibnamefont{Ramos}},
\bibinfo{author}{\bibfnamefont{S.}~\bibnamefont{Manley}},
\bibinfo{author}{\bibfnamefont{E.}~\bibnamefont{Pitard}},
\bibinfo{author}{\bibfnamefont{D. A.} \bibnamefont{Weitz}},
\bibinfo{author}{\bibfnamefont{E. E.} \bibnamefont{Pashkovski}},
\bibnamefont{and} \bibinfo{author}{\bibfnamefont{M.} \bibnamefont{Johansson}},
\bibinfo{journal}{Faraday Discussions} \textbf{\bibinfo{volume}{123}},
\bibinfo{pages}{237} (\bibinfo{year}{2003}).

\bibitem{bouchaud2001} J. P. Bouchaud and E. Pitard, Eur. Phys. J. E,  \textbf{6}, 231, 2001.

\bibitem{bouchaud90} J.-P. Bouchaud and A. Georges, Physics Reports, \textbf{195}, 127, 1990.

\bibitem{ruzicka11} B. Ruzicka, and E. Zaccarelli, Soft Matter,  \textbf{7}, 11551, 2011.

\bibitem{madsen10} A. Madsen, R. L. Leheny, H. Guo, M. Sprung, and O. Czakkel, New Journal of Physics,  \textbf{12}, 055001, 2010.

\bibitem{baret2002} J.-C. Baret, D. Vandembroucq and S. Roux, Phys. Rev. Lett.,  \textbf{89}, 195506, 2002.

\bibitem{PicardPRE2005} G. Picard, A. Ajdari, F. Lequeux and L. Bocquet, Phys. Rev. E,  \textbf{71}, 010501, 2005.

\bibitem{homer2009} E. R. Homer and C. A. Schuh, Acta Materialia,  \textbf{57}, 2823, 2009.

\bibitem{barrat2014} J.-L. Barrat, K. Martens, personal communication (2014).



\end{thebibliography}

\begin{thebibliography}{33}
\expandafter\ifx\csname natexlab\endcsname\relax\def\natexlab#1{#1}\fi
\expandafter\ifx\csname bibnamefont\endcsname\relax
  \def\bibnamefont#1{#1}\fi
\expandafter\ifx\csname bibfnamefont\endcsname\relax
  \def\bibfnamefont#1{#1}\fi
\expandafter\ifx\csname citenamefont\endcsname\relax
  \def\citenamefont#1{#1}\fi
\expandafter\ifx\csname url\endcsname\relax
  \def\url#1{\texttt{#1}}\fi
\expandafter\ifx\csname urlprefix\endcsname\relax\def\urlprefix{URL }\fi
\providecommand{\bibinfo}[2]{#2}
\providecommand{\eprint}[2][]{\url{#2}}

\bibitem[{\citenamefont{Hebraud}(1997)\citenamefont{Hebraud}}]{Hebraud1997}
\bibinfo{author}{\bibfnamefont{P.}~\bibnamefont{H\'{e}braud}},
\bibinfo{author}{\bibfnamefont{F.}~\bibnamefont{Lequeux}},
\bibinfo{author}{\bibfnamefont{J. P.}~\bibnamefont{Munch}},
\bibnamefont{and} \bibinfo{author}{\bibfnamefont{D. J.} \bibnamefont{Pine}},
\bibinfo{journal}{Phys. Rev. Lett.} \textbf{\bibinfo{volume}{78}},
\bibinfo{pages}{4657} (\bibinfo{year}{1997}).


\bibitem{PetekidisPRE2002} G. Petekidis, A. Moussaid and P. N. Pusey, Phys. Rev. E,  \textbf{66}, 051402, 2002.

\bibitem{Berne1976} B. J. Berne and R. Pecora, \textit{Dynamic Light Scattering}, Wiley, New York, 1976

\bibitem{LucaFaraday2003} L. Cipelletti, L. Ramos, S. Manley, E. Pitard, D. A. Weitz, E. E. Pashkovski and M. Johansson, Faraday Discuss.,  \textbf{123}, 237, 2003.

\bibitem{bouchaud90} J.-P. Bouchaud and A. Georges, Physics Reports, \textbf{195}, 127, 1990.

\end{thebibliography}
\end{document}